# THE ROLE OF QUALITY ASSURANCE IN SOFTWARE DEVELOPMENT PROJECTS:PROJECT FAILURES AND BUSINESS PERFORMANCE

**Ahmed Mateen**[*]

**Muhammad Jahanzaib***

**Nayyar Iqbal***

**Abstract**

In the product business still battles with the hard assignment of creating programming applications that meet quality gauges, and spending limitations. The requirement for programming to be without mistake remained a test to the IT business. Hence, the basic role of this study is to answer why officials hesitant to apportion assets to quality confirmation (QA) process amid the procedure of the framework advancement life cycle (SDLC)? This exploration utilized a quantitative study outline to research to what degree the inclusion of QA amid the SDLC procedure diminished programming venture disappointments. The information will be investigate utilizing inductive techniques and was expected to be summed up to the whole IT programming improvement populace. The exploration showed a feeling that incorporation of QA in all periods of SDLC was a great marvel. To gather the essential information, a proficient study will be led through survey structure will be utilized to take the criticisms and perspectives from various programming houses and industry specialists. The last stage will be the elucidating factual investigation by utilizing measurable techniques. Matlab programming will be utilized to gauge the fluctuation and execution by contrasting the execution parameters. As a consequence of this examination, the odds of the venture disappointment amid undertaking advancement will be declines and the business execution of the tasks will be expansions.

[*] **Department of Computer Science, University of Agriculture, Faisalabad**

## 1. Introduction

Software engineering is a field of computer science deals with designing, implementation and maintenance of computer systems. It covers all the technical and social aspects of building software systems, supervising development teams, scheduling and budgeting for the whole project. With the evolution of technology, this field became more challenging and demanding. Software engineers provide reliable and efficient software solutions for complex and real time problems. One of the well-known techniques is component based software engineering (CBSE). It deals with reuse-based approach to define required components, integrate them into systems. This development method is very different from other methods to develop systems in which the system is developed from scratch. In CBD, commercially available components are reused by different developers, using different languages, tools and technologies and then integrate those selected components into targeted software system. This approach is more efficient and time saving as compared to other traditional approaches but require a lot of effort to assure quality. In the age of growing technology and increasing the demand of more reliable and real time software, the complexity and the utilization of cost, time, and technical resources has been increased. On the other hand the risk of quality and move to new technology has also been increased (Rawashdeh and Matalkah, 2006).

Data systems/information technological innovation (IS/IT) with today's complicated in addition to dynamic environment is vital and development features carried on to grow; however, THE ITEM confronts the challenge of tips on how to properly build it is product or service with recognized techniques. Fees don't end result merely through producing in addition to repairing problems; a higher quantity of charges are derived from making certain excellent goods are made. The particular Team of Marketing, Economics, in addition to Stats Management within a analyze focused on on-line of application development, exhibited which started to advertise the standard Peace of mind career by way of spreading in addition to advancement of high specialized standards" Investment decision with application development increased through $82 billion with 1995 to $200 billion (Kaur and Mann, 2010).

Good quality peace of mind (QA) in this examine was described as any functionality together with obligation in order that software complies with its planned requirements—functional, date, personnel, spending budget, and many others. The same functionality bundled was software assessment. It turned out made to prove which a technique may satisfy its qualification. Interchangeably, QA in addition to software assessment was utilized since that examine discovered his or her inclusion within the technique advancement living routine (SDLC). Errors discovered earlier within the layout process could have rippling side effects, in addition to most of these glitches were being pricey in addition to difficult to improve immediately after software challenge completion (Rizwan and Shaukat, 2007).

Software package venture breakdowns could possibly be attributed to that which was claimed because issues in between enterprise also it management. Business management have also been allocating much less resources while they focused more in economics as it management devoted to techie advancement along with available technologies (Dixit and Saxena, 2011).

Ewusi-mensah brought up that 7% of it anticipates fulfilled are by and large 189% more than assets at yet another cost of $59 million. In addition, guided at the need to join individuals inside the product bundle advancement exertion by basically examining reactions which reminded THIS market, particularly this profession fields of programming bundle designing, to better understand this motivation behind people inside the reception procedure, alongside exactly how this connected with outlining after friendly logical orders sorts. There was obviously need to enhance programming bundle improvement with systems; all things being equal, these procedures are not a panacea for all product bundle advancement issues. The work of methods was viewed as an idealistic way towards getting great quality programming bundle stock (Ewusi-mensah, 2012).

The necessity with regard to software program to get error-free stayed challenging towards the idea sector. Benefit regarding accomplishment regarding software program improvement, that counted in efficiency, excellent, and also timeliness, except designed effectively to do it's preferred operate, the aim of software program improvement has been conquered. There was clearly numerous sides towards root cause regarding software program project breakdowns with increased exposure of a pair of with the principal brings about specifically, inadequate organizing and also an absence of training, especially in QA (Asif et. al., 2012).

## 2. Relatedwork

Bist et al. (2012) described that reusing the software is most important and comparatively innovative way in software engineering. A few reusability issues covering, allocation, arrangement, installation, operation, proceed and protection issues. Reusability takes some appearance to software development and provides innovative thoughts. The particular parts changes is utilized within software advancement procedure which usually boosts reusability method from various degrees for example from framework amount, structure amount along with modular design and style amount. Authors suggest a set of various methods produced will be recognizing the quality about reusable meets expectations. Yet the Different value of effort wishes to a chance to be finished on the structure for programming measurements with finding that reusability about COTS based software components.

Kumar and Singh (2012) stated that CBSE has become observable technique that speedy growth of system by means of a smaller amount resources and effort. CBSE gives initiative of reprocess and minimize those improvement cosset. Anyhow requesting of COTS components turns into additional complex the point when programmers are not gave for inner part plan to succession from claiming these COTS parts. Testing of black box mechanism is problem in the part of CBSE. CBSD not offer the combined Frameworks Similarly as black box trying for reusable segments. Testing of black box system will be issue in the and only CBSE. The prearranged content overview will comprise with respect to universal diaries made from multi-phase done combination system. They utilized equitably a little number from claiming testing instruments on like these testing measures. They additionally computed from claiming introduced trying methods Also its model for black box CBSD. They likewise segmentation resolved alongside trying about COTS segments manufactured clinched alongside trying. It might have been beneficial dealing with over states for research, thoughtful those measures of components, association and similarity for one another. The general trying about components, testing with respect to run-time also they utilized distinctive procedure with settle on test-cases for evolution segments.

Mantyla and Petersen (2012) incresed the volume of folks taking care of quality guarantee (QA) responsibilities, age. g., testimonials and also examining, boosts the volume of blemishes found - just about all increases the complete hard work unless hard work is usually manipulated having repaired hard work financial constraints. Each of our investigation investigates exactly how QA responsibilities need to be put together about two variables, when i. age. Period and also number of people. Most of us specify anseo problem to be able to reply this particular question. To be a central component the particular seo problem we examine and also identify exactly how recognition probability defect need to be modeled as being a purpose of your energy. Most of us apply the particular supplements utilised in madness with the seo problem to be able to empirical defect info of an research earlier performed having college or university individuals. The final results display that the maximum range of the volume of folks depends upon the exact defect recognition possibilities with the individual blemishes as time passes, but about the length of your energy funds. Future do the job may focus on generalizing the particular seo problem with a more substantial group of variables, as well as not merely job period and also quantity of folks but practical knowledge and also familiarity with the particular employees concerned, and also techniques and also instruments put on when performing some sort of QA job.

Nautiyal et al. (2012) discussed that moving in the direction of CBSE to create on the vital themes to developed function by maximum quality appropriate COTS components with well-define software erection. There would abundant sorts of models accessible for standard product improvement act which will be model about X, model from claiming Y and tie model yet all the toward those present days CBSE is using on large scale very speedily. They recommended a premeditated in collection components existence phase model which consisting on testing of components or verification as a hysterically course in each segment. They projected model name as Elite Life Cycle (ELCM) for CBD functioning as like predictable models. The key features from claiming provided for model will be reusability for throughout product progression, develop and process those re-usable segments exceptionally of service in obliging programming undertakings. Reusability may be a perfect work of art method throughout the software development. Which could a chance to be used to arrive at toward chiefly selecting re-usable parts etcetera re-builds them. Each programming improvement life cycle models need their careful repayment and difficulty.

Mantyla and Petersen (2012) incresed the volume of folks taking care of quality guarantee (QA) responsibilities, age. g., testimonials and also examining, boosts the volume of blemishes found - just about all increases the complete hard work unless hard work is usually manipulated having repaired hard work financial constraints. Each of our investigation investigates exactly how QA responsibilities need to be put together about two variables, when i. age. Period and also number of people. Most of us specify anseo problem to be able to reply this particular question. To be a central component the particular seo problem we examine and also identify exactly how recognition probability defect need to be modeled as being a purpose of your energy. Most of us apply the particular supplements utilised in madness with the seo problem to be able to empirical defect info of an research earlier performed having college or university individuals. The final results display that the maximum range of the volume of folks depends upon the exact defect recognition possibilities with the individual blemishes as time passes, but about the length of your energy funds. Future do the job may focus on generalizing the particular seo problem with a more substantial group of variables, as well as not merely job period and also quantity of folks but practical knowledge and also familiarity with the particular employees concerned, and also techniques and also instruments put on when performing some sort of QA job.

Iqbal et al. (2013) identified in which advancement procedure which usually boosts reusability method from various degrees for example from framework amount, structure amount along with modular design and style amount. The particular advancement means of CBSE modifies the particular reusability method into a pair of various solutions, generation-based method along with composition-based method, which can be quite helpful when coding parts tend to be reused. Your overall performance metrics for software program pattern and also software program undertaking administration. Course of action improvement systems are generally elaborated in carried out quality warranty and also mentioned agile software program progress techniques and also analyzed issues pertaining to prerequisite engineering techniques. Furthermore, it describes non-functional demands for software program maintainability.

Lahon and Sharma (2014) described that CBSE guarantee model to product improvement yet it will be still on accomplish development as far as result life cycle terminology. There are different challenges which necessitate to be addressed. The key confront in this domain is getting a finest conversion procedure from requirements to components and then system to components. Similarly, as necessities deviate from customer to customer. It gets convoluted will distinguish what's more erect those part faultlessly will satisfy the prerequisite. Along these lines marginally change will be needed with reusing part. Further challenges in this region includes the issue of consistency in terms of deliverables component, time and exertion in developing components. Ambiguous and vague requirements, component maintenance usability and reusability also facing the challenge.

Mukherjee et al. (2013) described that programming cost estimation is the procedure of anticipating the expense regarding endeavors required to build up a product item. Various elements add to general cost estimation procedure of programming however variables, for example, programming size and its intricacy influences significantly to the exertion programming advancement precisely. Numerous models and measurements that have been proposed throughout the most recent 30 years in the writing keeping in mind the end goal to lessen programming advancement cost. This paper compresses the distinctive models and measurements. Additionally, gives a review of programming cost estimation instruments which is key for evaluating. As the volume and multifaceted nature of programming application are consistently expanding for that cost estimation turns out to be exceptionally exertion concentrated assignment.

Sandeep et al. (2014) exposed that component qualifications may be the system that includes progressively a part outsourcing agreement also auditing the administration supplier presentation. Choose on the exact parts done understanding of the state to both. . The particular parts changes is utilized within software advancement procedure which usually boosts reusability method from various degrees for example from framework amount, structure amount along with modular design and style amount.

### 3. Mathedology
#### *3.1 Conceptual Framework*
In the reasonable structure, the advancement of data frameworks included recognizing what to make furthermore how to make it. It's demonstrated that necessities get-together were frequently distinguished as the most troublesome piece of bringing a data framework into reality. The framework examiner's capacity to create precise, finish, and clear data necessities was vital for fruitful frameworks building. Prerequisites assurance amid data frameworks conveyance was a complex authoritative attempt, the hypothetical structure for the study identified with the administration hypotheses from IT, and hierarchical execution as far as the impacts of QA and IT on authoritative execution and the part of QA in programming improvement usage.

In the reasonable system, quickly after the fulfillment of prerequisites assembling, all partners ought to audit them before the following stride, which is presentation to administration for spending endorsement.

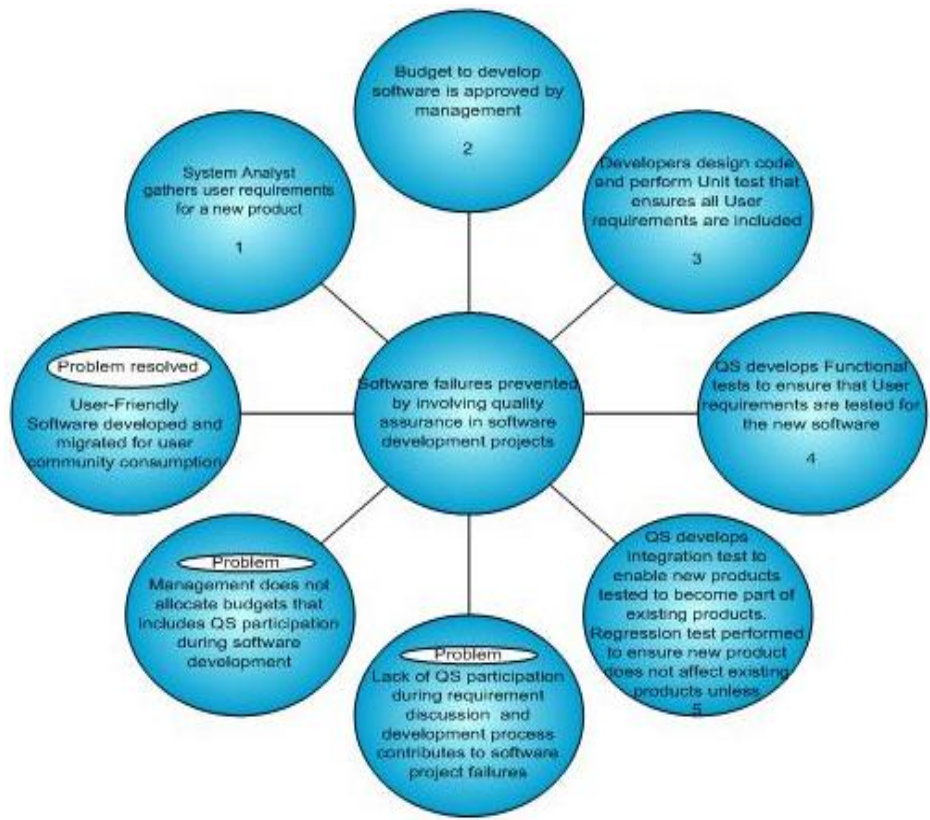

**Conceptual Framework**

*3.2 Assessment of current literature*

The necessity with regard to software program to get error-free stayed challenging towards the idea sector. Benefit regarding accomplishment regarding software program improvement, that counted in efficiency, excellent, and also timeliness, except designed effectively to do it's preferred operate, the aim of software program improvement has been conquered. There were clearly numerous sides towards root cause regarding software program project breakdowns with increased exposure of a pair of with the principal brings about specifically, inadequate organizing and also an absence of training, especially in QA (Asif et. al., 2012).

The computer software failure taken place as soon as an item of it still did not accomplish seeing that end users expected and also predicted. "The U. Azines.financial system spent virtually $60 thousand each year on account of computer software problems cost". Having talked about these kind of remarkable seems to lose towards U. Azines. financial system on account of computer software problems, we have a must distinguish the best way to get rid of or maybe minimize these kind of deficits appreciably.

To collect the primary data, an efficient survey will be conducted regarding effectiveness, performance and QA evaluation of SDLC process reduces software project failures. Furthermore, a questionnaire form will be use to take the feedbacks and views from different software houses and industry experts. This may valuable inside the supplement connected with QA to all stages of development connected with SDLC was a very good happening.

Different views will be collected from different organizations and experts. The main objective of questionnaire will be investigating and measure the immersion of SQ reduction factors. The managers, Software engineers and software quality related persons would respondents of questionnaire. The final phase will be the descriptive statistical analysis by using statistical methods. Matlab software will be used to measure the variance and performance by comparing the performance parameters.

## 4. Results

The demographic properties of the examination are outlined in it. Figure 1 shows the scattering of demographic investigation (e. g. sexual introduction, bundle, length of experience and pay) in related association.

Figure 1 demonstrate that male workers contain the larger part of faculty: 89. 00 for each dollar of this organization intended to utilize dexterous procedure. In any case, there were also 36. 00 for every dollar women, making up an also low degree of agents. It must be indicated that men work in wide territories of association and as show in figure 2 around 88. 80 % of delegate are given and only 11. 20 of these single. This figures exhibit this typical age social event of delegates are above 25 years old and one of principal goal of the association is contracting experienced, and submitted.

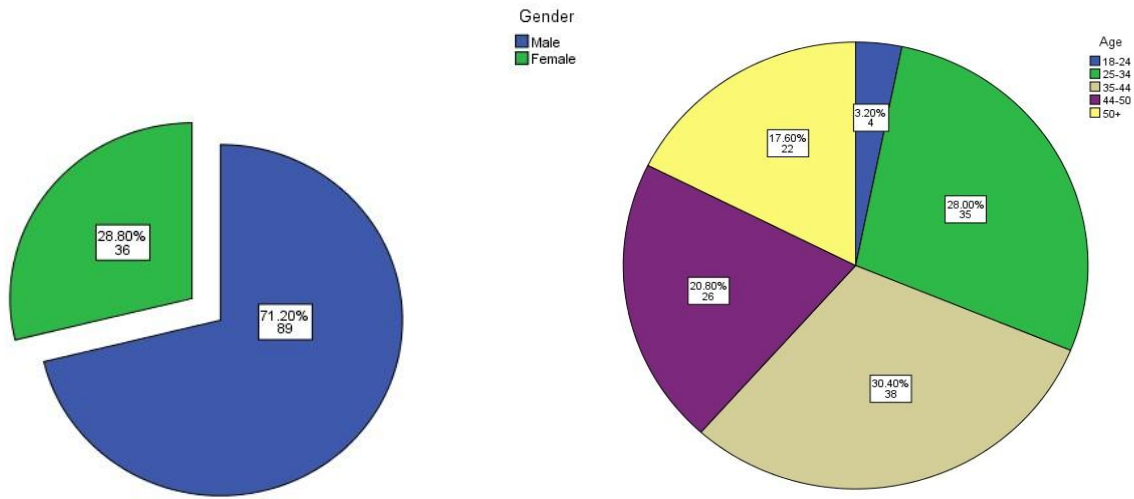

Figure 1.Age and sex information

Figure 2 displays bunch thirty-five ~ 44 by 40. 40 percent and the second the first is in the age 25 ~ 34 with 28. 00 for every penny.In any case, the workers in the scope of 44 ~ 40 with 20. 60 each penny are the third most noteworthy gathering, while representatives with between 18 ~24 would be the littlest gathering with just 3. 20 each penny and 50+ age gathering are just 18. 60 for every penny.This sort of determination of worker between a quarter century 50 impact wok exercises in deft procedure experience.

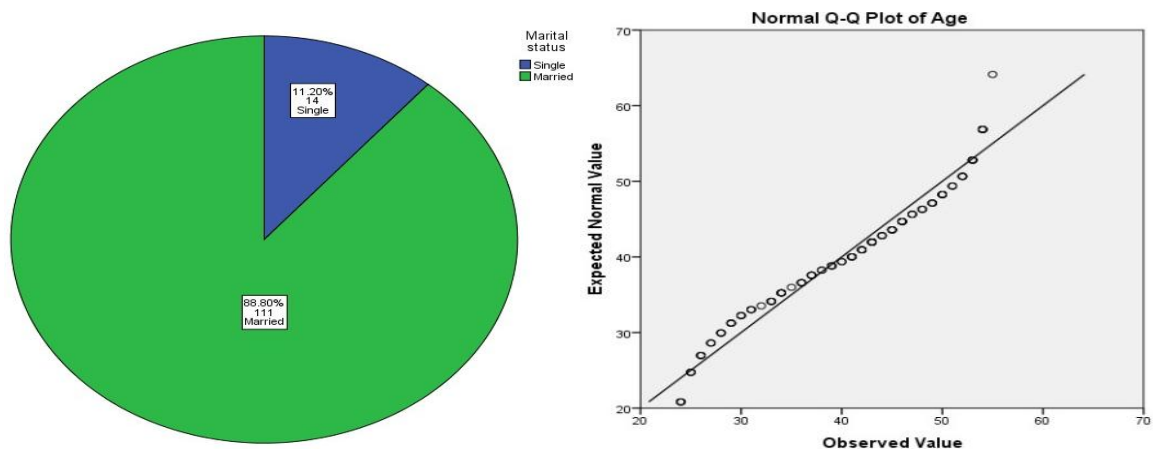

Figure 2.Gender married status

Figure 3 exhibits the work practices course of agents in this association. Specialist with between 1-5 and 6-10 years' experience built up the two most basic get-togethers of staff in the association. Together they included more than 60% of all people in the organization. The laborers with 1-5 years' experience spoke to around 50 for each penny of all agents. So also, work power with 6-10 years' experience included around 18 every penny of the total. In examination, the delegates with 11-15 years' experience have around 17. 60 every penny and between 16-20 years practically 15.20 for every penny. It can be seen than the agents with more than 20 years' experience involve only 0 for every dollar the organization.

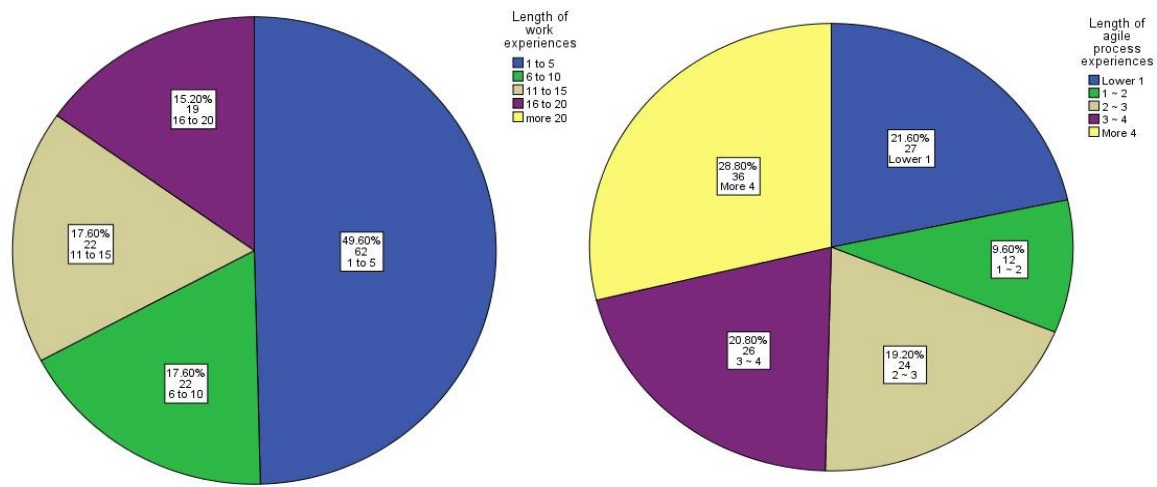

Figure 3.Worth of experience

## 5. Conclusion

This investigation addresses that fuse of QA in all times of SDLC was an awesome wonder. Much the same as each other thing intension, there were still opponents to the gathering of the thought. 72% of the diagram masses was of the conclusion that QA should be exhibited in all times of the SDLC. The remaining 28% fell into an alternate evaluations groups. In a response from an eye to eye chat with, one part quickly related the unsteady economy as an overall constrain against incorporating QA in all times of the SDLC. Diverse examples that affected the consolidation of QA in all times of the SDLC included

(a) the procedure use an excessive amount of time

(b) Inconsistent approachs

(c) The tasks finish in as far as possible,

(d) Unclear necessities cause the disappointment of the activities in the majority of the circumstances.

## 6. Future work

Case in point, correspondence may influence data trade than on learning creation and support, as correspondence in programming planning is much of the time related to learning trade. In addition, legitimate society is a wide thought that has various estimations.

Future examination can dissect particular operational of definitive society, furthermore social complexities between individual planners.

This examination study set out to use study data to explore the fundamental accomplishment or disillusionment segments of composed programming change wanders using quantitative strategies. The data assembled from 125 specialists who contribute in association gave enough trial information to authentic examination to meet up at different conclusions.